\documentclass[german,english,aps,color,preprint,memoir]{revtex4}
\usepackage[latin9]{inputenc}
\usepackage{float}
\usepackage{bm}
\usepackage{amsmath}

\usepackage{amssymb}
\usepackage{graphicx}
\usepackage{esint}
\usepackage{colortbl}

\usepackage[T1]{fontenc}

\makeatletter

\medskipamount 

\makeatother

\usepackage{babel}
\begin{document}

\preprint{This line only printed with preprint option}

\vspace*{1cm}

\title{Wavefunctions for large electronic systems}

\author{Peter Fulde}
\email{fulde@pks.mpg.de}
\affiliation{Max-Planck-Institut f\"ur Physik komplexer Systeme, N\"othnitzer Stra\ss e 38, 01187 Dresden, Germany}
\date{\today}

\begin{abstract}
\noindent Wavefunctions for large electron numbers suffer from an exponential growth of the Hilbert space which is required for their description. In fact, as pointed out by W. Kohn, for electron numbers $N > N_0$ where $N_0 \approx 10^3$ they become meaningless (exponential wall problem). Nevertheless, despite of the enormous successes of density functional theory, one would also like to develop electronic structure calculations for large systems based on wavefunctions. This is possible if one defines the latter in Liouville space with a cumulant metric rather than in Hilbert space. The cluster expansion of the free energy of a classical monoatomic gas makes it plausible that cumulants are a proper tool for electronic structure calculations.

\vspace*{2cm}

\begin{center} 
{\it Invited chapter in\\`Many-body approaches at different scales - A tribute to N. H. March\\on the occasion of his 90th birthday',\\edited by G. G. N. Angilella and C. Amovilli (New York : Springer, to be published)}
\end{center}
\end{abstract}
\maketitle


Electronic structure calculations, in particular for large systems, are one of the most active and challenging fields in condensed matter physics and quantum chemistry. This remains true despite the fact that the field has almost exploded during the last thirty years. Naively one would think that in an electronic structure calculation one is aiming to determine the many-body wavefunction of the interacting electron system and to derive from it physical properties of interest. This was the path taken when shortly after the rules for dealing with quantum mechanical systems were formulated by Heisenberg \cite{Heisenberg25} and Schr\"odinger \cite{Schroedinger26} these were applied by Hund, Mullikan, Heitler, London and others in order to study chemical binding, thereby beginning with the $H_2$ molecule. Since then the sizes of the quantum chemical systems for which the electronic structures were studied grew continuously. At present, even electrons in molecules with hundreds of atoms have been successfully treated (see, e.g., \cite{Werner12,Liakos15}), as well as in solids with periodic lattices (see, e.g., \cite{Manby11}).

However, with increasing electron numbers $N$ the dimension of the Hilbert space spanned by the different electronic configurations increases exponentially with $N$. This led Walter Kohn to the statement \cite{Kohn99} that the many-electron wavefunction for a system of more than $N \approx 10^3$ electrons is not a legitimate scientific concept anymore. He referred hereby to wavefunctions expressed in Hilbert space and required that for a legitimate scientific concept two conditions have to be fulfilled: it should be possible to calculate the wavefunction with sufficient accuracy and it should be possible to represent it numerically sufficiently well. Because of the exponential growth of the Hilbert space none of the two conditions can be satisfied when $N \gtrsim 10^3$. Any approximation $\psi_{\rm calc}$ to the exact ground-state wavefunction $\psi_0$ will have an overlap with the latter of order $| \langle \psi_0 | \psi_{\rm calc} \rangle |^2 = (1 - \epsilon)^N$, which is zero for all purposes, if $N \to \infty$. Here $\epsilon$ is the minimum error one has to deal with when approximations for the description of an electron are being made. A similar argument applies to the second condition. When it needs $m$ bits to describe a single interacting electron, the total number of bits is $m^N$ in order to describe the full electron system.

The exponential growth of Hilbert space is called the exponential wall (EW) problem. The simplest way to get around it is by making use of density-functional theory (DFT), developed by Hohenberg, Kohn and Sham \cite{Hohenberg64,Kohn65} (for extension see e.g., \cite{March99}). Here all degrees of freedom of the electronic system are integrated out, except for the density. No statements are required about the many-electron wavefunctions. The strength of DFT is based on this feature.

Another way of avoiding the EW problem is by reducing the electron Hamiltonian $H$ to its self-consistent field (SCF) part $H_{\rm SCF}$. This simplifies the problem to a single-electron one with a potential  which has to be determined self-consistently. The ground-state wavefunction is given in this case by a single Slater determinant or configuration. Correlation effects are hereby completely neglected and therefore results for various physical quantities are usually of low quality.

Although DFT has revolutionized the field of electronic structure calculations, and can claim phantastic successes, it contains also weaknesses. A general one is that its results depend on the chosen exchange-correlation potential and any approximation to it is essentially uncontrolled. This leads to problems when electronic correlations are strong or when one is dealing with dispersive electron interactions.

For the above reasons it seems worthwhile to develop in parallel to DFT also electronic structure calculations based on wavefunctions. This approach is favoured by the accuracy of quantum chemical techniques in cases when they have been applied. The question therefore is: does Kohn's correct argument about the inadequacy of wavefunctions for large systems prevent us from doing such calculations on firm theoretical grounds? In short, the answer is no! However, one has to give up characterizing the many-electron wavefunction in Hilbert space. Instead it has to be characterized in Liouville or operator space. The reason is not difficult to see. Consider $A$ very weakly interacting atoms with $N_A \geq 2$ electrons each. The dimension of Hilbert space for the ground state of the electrons on a single atom is of order $d^{N_A}$ with $d \geq 2$. Yet the dimension required for a description of the whole system is $d^{A \cdot N_A}$ despite the fact that in the limit of weak coupling between electrons on different atoms the wavefunction does not contain any additional information compared with the one obtained from a single atom. Thus disconnected correlations are responsible for the EW problem. This suggests to eliminate all disconnected contributions to the wavefunction in order to free oneself from the EW problem. This is easily done with the help of cumulants.

\section*{Use of cumulants}

Cumulants of matrix elements eliminate factorizable contributions to it. In the simplest case the cumulant, denoted by $c$, of a product of two operators $A_1 A_2$ sandwiched between to vectors $\phi_1$ and $\phi_2$ in Hilbert space with $\langle \phi_1 \mid \phi_2 \rangle \neq 0$ is 

\begin{equation}
\label{Eq1}
\langle \phi_1 | A_1 A_2 | \phi_2 \rangle^c = \frac{\langle \phi_1 | A_1 A_2 | \phi_2 \rangle}{\langle \phi_1 | \phi_2 \rangle} - \frac{\langle \phi_1 | A_1 | \phi_2 \rangle \langle \phi_1 | A_2 | \phi_2 \rangle}{(\langle \phi_1 | \phi_2 \rangle)^2} ~~.
\end{equation}\\
Note that a replacement of $| \phi_2 \rangle$ by $\alpha | \phi_2 \rangle$ with $\alpha \neq 0$ leaves the cumulant unchanged. General rules for cumulants are found in the literature, see e.g., \cite{Kladko98,Fulde95_12}. They were first applied in statistical physics \cite{Mayer40} when dealing with the classical imperfect gas and later pioneered by Kubo \cite{Kubo62} in quantum statistical mechanics. In practice a cumulant implies taking only connected contractions of operators into account when a matrix element or expectation value is evaluated. With this in mind we divide the electronic Hamiltonian $H$ into $H = H_0 + H_1$ so that the ground state of $H_0$, i.e., $| \Phi_0 \rangle$ can be easily calculated. Often $H_0$ will be $H_{\rm SCF}$, yet it can be also, e.g., the Kohn-Sham Hamiltonian $H_{\rm KS}$. The remaining part $H_1$ contains the residual interactions. We define $| \Phi_0 \rangle$ as the vacuum state so that $H_1$ generates fluctuations, i.e., vacuum fluctuations on it. Next we consider a matrix element of an arbitrary operator $A$ with respect to $| \Phi_0 \rangle$, i.e., $\langle \Phi_0 | A | \Phi_0 \rangle = \langle \Phi_0 | A | \Phi_0 \rangle^c$. By a sequence of infinitesimal transformations in Hilbert space, we transform the state $| \Phi_0 \rangle$ on the right, into the exact ground state $| \psi_0 \rangle$ of $H$. Then the above expression transform into
\begin{equation}
\label{Eq2}
\langle \Phi_0 | A | \psi_0 \rangle^c = \langle \Phi_0 | A \Omega | \Phi_0 \rangle^c\end{equation}
with $\Omega = 1 + S$ and $S$ denoting the sum of the infinitesimal increments in the transformation. For a more accurate derivation of (\ref{Eq2}) see \cite{Kladko98}. Note that $\Omega$ is not unique since many different paths in Hilbert space may lead from $| \Phi_0 \rangle$ to $| \psi_0 \rangle$. Yet, the cumulant remains unchanged by these differences.

In quantum mechanics the operator which transforms the ground state of an unperturbed system (here $| \Phi_0 \rangle$) into the one of the perturbed system (here $| \psi_0 \rangle$) is called M{\o}ller- or wave operator, i.e., $| \psi_0 \rangle = \tilde{\Omega} | \Phi_0 \rangle$. Therefore we call $\Omega$ in (\ref{Eq2}) a cumulant wave operator and $S$ in $\Omega = 1+S$ a cumulant scattering operator. Eq. (\ref{Eq2}) suggest to introduce the following metric in Liouville space.
\begin{equation}
\label{Eq3}
(A | B) = \langle\Phi_0 | A^+ B | \Phi_0 \rangle^c~~.
\end{equation}
The metric is not a scalar product since it may vanish or even became negative.

\section*{Cumulant scattering matrix}

The exponential wall problem does not exist if we define the many-electron wavefunction not by a vector in Hilbert space but by the vector $| \Omega )$ in Liouville space. An exponentially small overlap of $\langle \psi_0 | \psi_{\rm cal} \rangle$ is harmless since the cumulant, e.g., in (\ref{Eq1}) remains unchanged when $\phi_1$ is replaced by $\alpha \phi_1$ with $\alpha \neq 0$. When $| \psi_{\rm cal} \rangle$ instead of $| \psi_0 \rangle$ is considered the only effect is a change of $|S)$ to $|S+\delta S)$. Also a numerical representation of $| \Omega )$ or $|S)$ poses no problems. To see this we decompose $|S)$ into increments
\begin{equation}
\label{Eq4}
| S ) = \left| \sum_I S_I +  \sum_{\langle IJ \rangle} \delta S_{IJ} + \sum_{\langle IJK \rangle} \delta S_{IJK} + \dots \right) 
\end{equation}
where $I, J, K$ etc are site indices. Here $\delta S_{IJ} = S_{IJ} - S_I - S_J$ etc. When $S_I$ is calculated all electrons in $|\Phi_0 \rangle$ are kept frozen except those in orbitals centered at site $I$. The procedure is similar when $S_{IJ}$, $S_{IJK}$ etc are calculated. In each case only a small number of electrons is involved and therefore the different increments of $|S)$ can be documented without problem. Thus the EW problem has been eliminated.

\begin{figure}[h]
\includegraphics[width=0.6\textwidth]{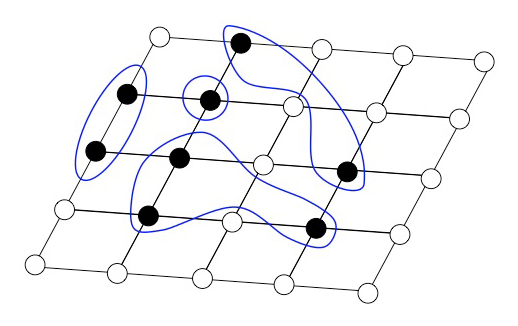}
\caption{Examples of different vacuum fluctuations $S_I$, $S_{KL}$, $S_{MNT}$ contributing to $|S)$.}
\label{fig:01}
\end{figure}

\begin{figure}[h]
\includegraphics[width=0.6\textwidth]{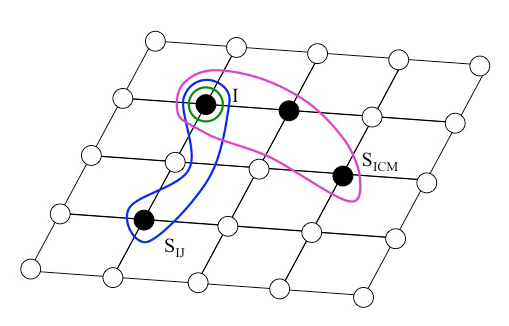}
\caption{Vacuum fluctuations contributing to $E_{\rm corr} (I)$. Different colours refer to connected vacuum fluctuations involving electrons on different numbers of sites.}
\label{fig:02}
\end{figure}

Let us consider $|\Phi_0 \rangle$ as the vacuum state of the system. The operators $S_I$, $S_{IJ}$ etc can be thought of generating vacuum fluctuations. They modify the energy $E_0$ of the system according to
\begin{equation}
\label{Eq5}
E_0 = \langle \Phi_0 | H \Omega | \Phi_0 \rangle^c = (H | \Omega)~~.
\end{equation}
When $H_0 = H_{\rm SCF}$ the correlation energy is simply
\begin{eqnarray}
\label{Eq6}
E_{\rm corr} & = & \left( H_1 | \Omega \right) \nonumber \\[2ex]
& = & \left(H_1 | S \right)~~.
\end{eqnarray}
Because of the cumulant metric only connected vacuum fluctuations contribute to $E_{\rm corr}$. This is depicted in Fig. \ref{fig:01}. Note that $I$ and $J$ need not be nearest neighbors. For a given site $I$ the correlation contributions add up to
\begin{equation}
\label{Eq7}
( H_1 | S_I + \sum\limits_{J \neq I} \delta S_{IJ} + \dots ) = E_{\rm corr} (I)
\end{equation}
and $E_{\rm corr} = \sum\limits_I E_{\rm corr} (I)$. This is indicated in Fig. \ref{fig:02}. The $\delta S_{IJ}$, $\delta S_{IJK}$ etcetera decrease rapidly with increasing number of subscripts. They also decrease rapidly with increasing distances of the sites $I$, $J$ etc. The slowest decrease is taking place for $\delta S_{IJ}$ when $|{\bf R}_I - {\bf R}_J|$ increases where ${\bf R}_I$ denotes the position of site $I$. Except in the vicinity of an electronic phase transition the decrease of $(H_1 | \delta S_{IJ})$ with $|{\bf R}_I - {\bf R}_J| \to \infty$ is exponential and defines a characteristic correlation length, e.g., in a solid.

The behavior of $(H_1 | \delta S_{IJ})$ for a given system in this limit tells us whether or not an area law \cite{Eisert10} is holding. We want to point out that the theory presented here has been applied to a number of solids, mainly to semiconductors and insulators. Metallic systems \cite{Paulus04,PaulusB04} require additional comments since their occupied Wannier functions are not well localized \cite{Kohn59}. Yet, this does not pose a principle problem. A review of the different results obtained is found in \cite{Manby11,Fulde95_12,Paulus06} although the way is not always obvious in which the computations described there relate to the computational scheme outlined here.

\section*{Classical imperfect gas}

It is instructive to compare the definition of a wavefunction in Liouville space and the computation of the correlation energy with the one of the temperature $T$ dependent free energy $F(T)$ of a classical imperfect (real) monatomic gas. The limit $F(T=0)$ yields the classical analogue of the electronic ground-state energy. Let $U = \sum\limits_{i > j} \phi_{ij}$ denote the potential energy of the system where $\phi_{ij}$ denotes the pair interactions of the gas particles. With $f_{ij} = {\rm exp} (- \beta \phi_{ij}) -1$ and $\beta = (k_B T)^{-1}$ we can write
\begin{equation}
\label{Eq8}
e^{- \beta U} = \prod_{i > j} e^{- \beta \phi_{ij}} = \prod_{i > j} \left( 1 + f_{ij}\right)~~.
\end{equation}
The partition function $Z$ is the product $Z = Z_{id} \cdot Z_U$ of the one for the ideal gas $Z_{id}$ and
\begin{equation}
\label{Eq9}
Z_U = \frac{1}{V^N} \int d{\bf r}_1 d{\bf r}_2 \dots dr_N e^{- \beta U ({\bf r}_1, \dots, r_N)}~~.
\end{equation}
The integration is over the coordinates of the $N$ particles and extends over the full volume $V$ of the system. Thus the free energy $F(T) = F_{id} + F_U$ can be written in the form
\begin{eqnarray}
\label{Eq10}
F & = & F_{id} - k_B T \ln \frac{1}{V^N} \int d{\bf r}_1 \dots d{\bf r}_N e^{- \beta U ({\bf r}_1, \dots, r_N)} \nonumber\\
F_U & = & -k_B T \ln \left< \prod_{i > j} \left( 1 - f_{ij} \right) \right>
\end{eqnarray}
where $F_{id}$ is the free energy of the ideal gas.

An often used definition of cumulants is of the form
\begin{equation}
\label{Eq11}
\ln \langle e^{\lambda A} \rangle = \langle e^{\lambda A} - 1 \rangle^c
\end{equation}
and demonstrates how cumulants avoid dealing with the logarithm of averages. When applied to $F_U$ we can rewrite the latter as
\begin{equation}
\label{Eq12}
F_U = -k_B T \left< \sum_{i < j} f_{ij} + \sum_{i < j} \sum_{k < l \atop i,j \neq kl} f_{ij} f_{kl} + \dots \right>^c~~.
\end{equation}
Thus to the free energy of a real gas \cite{Mayer40} only {\it linked} pair interactions do contribute. This is sometimes referred to as Mayer's cluster expansion.

This well known classical results suggest strongly that also in quantum mechanics cumulants are a proper tool for calculating energies for large systems of interacting particles. It makes the choice of the metric in Liouville space rather obvious.

The question remains, how the present approach based on a description of wavefunctions in Liouville space with cumulant metric compares with other approaches avoiding the EW problem. A comparison with the Density Matrix Embedding Theory (DMET) \cite{KnizChan12} was recently worked out \cite{FuldeStoll}. A corresponding one for the Density Matrix Renormalization Group (DMRG) \cite{Schollwoeck05} as well as with tensor networks \cite{Verstraete08} in particular with Matrix Product States \cite{Oros14} is left for the future.

\section*{Conclusions}

Wavefunctions defined in Hilbert space are no longer meaningful when the electron number exceeds approximately $10^3$. The reason is the exponential growth of the dimensions of this space with electron number. But when instead wavefunctions are described in operator- or Liouville space, this problem is circumvented. The operators spanning Liouville space describe the quantum fluctuations out of a mean-field ground state configuration. These fluctuations must be linked in order to contribute to the wavefunction, similarly as interaction contributions to the free energy of a classical gas have to be linked. This is ensured by choosing a cumulant metric for the Liouville space.

\section*{Acknowledgement}

I would like to thank Hermann Stoll for many fruitful discussions and collaborations on subjects related to this article.

\end{document}